\documentstyle[preprint,epsfig,eqsecnum,aps]{revtex}
\newcommand{\bea}{\begin{eqnarray}}
\newcommand{\eea}{\end{eqnarray}}
\newcommand{\beq}{\begin{equation}}
\newcommand{\eeq}{\end{equation}}
\begin{document}
\preprint{\parbox{4cm}{\flushright CPPM 95-04\\ MZ-TH/95-28}}
\title{Improved variables for measuring the $\Lambda_{\mathrm{b}}$
polarization}
\author{C.Diaconu, M.Talby}
\address{Centre de Physique des Particules, Facult\'e des Sciences de Luminy,
IN2P3-CNRS, 13288 Marseille, France}
\author{J.G.K\"orner, D.Pirjol}
\address{Johannes Gutenberg-Universit\"at,
Institut f\"ur Physik (THEP), Staudingerweg 7,\\
D-55099 Mainz, Germany}
\date{\today}
\maketitle
\begin{abstract}
We discuss a few possible strategies for measuring the polarization of the
$\Lambda_{\mathrm{b}}$ baryons produced in
 $e^+e^-$-annihilation at the $\mathrm{Z}$ resonance
through their inclusive semileptonic decays. After reviewing the existing
methods, an extension  is proposed, based
on the ratio of the averages of the squared electron and neutrino energy,
including both perturbative and nonperturbative corrections. This variable
minimizes the statistical error on the $\Lambda_{\mathrm{b}}$ polarization,
 while
keeping the systematic theoretical errors at the level of 1-2 \%. A number of
other polarization-sensitive variables are also discussed,
such as averages of ratios
of the electron and neutrino energy and the distribution in the difference
of the electron and neutrino rapidities.
\end{abstract}
\narrowtext
\section{Introduction}
 It is a well-known fact that the bottom quarks produced in
$e^+e^-$-annihilation at the $\mathrm{Z}$ peak are predicted to be very
strongly
polarized. This can be understood by noting that at this energy the
production process is practically dominated by $\mathrm{Z}$-exchange,
 whose coupling to quarks
is proportional to $T_3-\sin^2\theta_W Q$. In the limiting case when
$\theta_W=0$, the right-handed quarks (for which $T_3=0$) decouple completely
from the neutral current and thus the produced quarks are purely
left-handed.
Increasing the value of the weak angle $\theta_W$ to its physical value
somewhat diminishes the effect (more for the $\mathrm{u}$-type
 than for the $\mathrm{d}$-type
quarks), such that for $\sin^2\theta_W$=0.23, a polarization degree
$P=-0.936$ is expected for the $\mathrm{d}$-type quarks.

  The $\mathrm{b}$-quark mass effects are suppressed
 by a factor of $m_{\mathrm{b}}^2/m_{\mathrm{Z}}^2$ and
are very small, less than one tenth of a percent.
The radiative corrections to this prediction have been computed to one-loop
order \cite{1,2,3} and have been found
to decrease $P$ slightly, by about 2\%. Also, the
polarization depends only weakly on the kinematical details of the
production process.

  Eventually the $\mathrm{b}$-quark is observed through its hadronization
products. The $b$ quark hadronization into mesons (even excited ones) deletes
any memory of the original quark polarization \cite{FP}.
On the other hand, if the quark ends up as a
 $\Lambda_{\mathrm{b}}$ baryon (in which
the light constituents combine to spin 0), the latter is expected
to retain a large part of the initial polarization of the quark \cite{CKPS}.

  A measurement of the polarization
 of the produced $\Lambda_{\mathrm{b}}$'s could
therefore help to test directly this prediction of the Standard Model
and obtain information about the details of the hadronization process.
A first measurement is already available \cite{PPE}, which gave an
intriguingly small value $P=-0.23^{+0.26}_{-0.23}$.
This measurement made use of the method proposed in \cite{BR}, based on the
ratio of the average electron and neutrino energies $y=\langle E_\ell
\rangle/\langle E_\nu\rangle$ in the laboratory frame.
This is just one of the many approaches which have been given which are
sensitive to $\Lambda_{\mathrm{b}}$ polarization effects, based on its
semileptonic decays to charmed hadrons.
Thus, in \cite{AM,M} the average lepton energy in the
laboratory frame has been suggested as a measure of the $\Lambda_{\mathrm{b}}$
polarization. Unfortunately, this quantity is rather sensitive to the
exact values of the quark masses, which are known only imprecisely.

   In contrast, the variable $y=\langle E_\ell\rangle/\langle E_\nu\rangle$
proposed in \cite{BR}
presents the advantages that the poorly known $\mathrm{b}$-quark mass drops out
and the dependence on the ratio $m_{\mathrm{c}}^2/m_{\mathrm{b}}^2$ is
compensated to a large extent.
Also, the dependence on the details of the fragmentation process
(the fragmentation function) disappears almost completely, a problem which
had been circumvented in \cite{AM,M} by a combined use of low-energy
experimental data. Recently, yet a different method has been put forward
in \cite{KK}.

  The purpose of the present paper is twofold: a) to give a detailed analysis
of the theoretical errors which affect an extraction from experiment of
the $\Lambda_{\mathrm{b}}$ polarization and b) to discuss a number of other
variables which can be used to measure the polarization. We find that the
theoretical uncertainty is dominated by the one in the ratio of the
quark masses $\rho=m_{\mathrm{c}}^2/m_{\mathrm{b}}^2$ and by the uncertainty
in the strong coupling $\alpha_s$.
In section 2 we discuss a generalization of the variable proposed in
\cite{BR}, the ratio of the moments of the electron and neutrino spectrum.
We argue that the ratio of the second moments minimizes the statistical
errors in the polarization, while being also insensitive to theoretical
uncertainties.
In section 3 two other polarization-dependent variables are proposed, the
averages of ratios of the electron and neutrino energies and the
decay distribution in the difference of the electron and neutrino
rapidities.


\section{Moments of the lepton spectra}

  The polarization of a $\Lambda_{\mathrm{b}}$ baryon affects directly the
spectra of the leptons ($e$ and $\nu$) produced in its semileptonic decays.
The qualitative nature of the change can be obtained by examining the angular
distribution of the leptons in the rest frame of the decay: the electrons
tend to be emitted antiparallel to the spin of the $\Lambda_{\mathrm{b}}$,
whereas the $\bar\nu$'s are emitted preferentially parallel to the spin.
In the laboratory system this translates into a larger number of electrons
being emitted collinear with the jet containing the decaying $\mathrm{b}$
quark (assuming negative polarization) and hence more energetic as compared
with the unpolarized case (and the opposite for neutrinos).
Consequently, the electron spectrum in the laboratory frame will be
hardened in the presence of negative $\Lambda_{\mathrm{b}}$ polarization,
contrary to the energy spectrum of the neutrinos, which is softened.

   It is more convenient to study the moments of the lepton spectra, rather
than the spectra themselves. The theory of the inclusive semileptonic decays
of heavy hadrons predicts that the lepton spectra are given, in a first
approximation, by the free-quark decay results. The corrections to these
results
are of order $1/m_{\mathrm{b}}^2$ \cite{CGG,Blok,MW} and can be parametrized
in terms of a few
matrix elements. We write the general form of the electron spectrum in the
rest frame of the decay as
\bea\label{1}
\frac{\mbox{d}\Gamma}{\mbox{d}x_e\mbox{d}\cos\theta_e} =
\frac{G_F^2|V_{cb}|^2}{24(2\pi)^3}m_{\mathrm{b}}^5
\left(F_e(x_e,\rho) + P\cos\theta_e J_e(x_e,\rho)\right)\,,
\eea
where $x_e=E_2/2m_{\mathrm{b}}$ is the reduced electron energy, $\theta_e$
is the angle
between the electron momentum and the $\Lambda_{\mathrm{b}}$ momentum,
$\rho=m_{\mathrm{c}}^2/m_{\mathrm{b}}^2$, and $P$ is the polarization.
In writing this we have assumed that
the $\Lambda_{\mathrm{b}}$ is longitudinally polarized, which is true to a
high degree
of accuracy\footnote{The transverse polarization, averaged over the
orientation angle with respect to the $e^+e^-$ beam, is predicted to be
less than 1\% in the Standard Model \cite{tr1,tr2}.}. Thus $\vec P$ is
antiparallel to the $\Lambda_{\mathrm{b}}$ momentum.

 To leading order in $1/m_{\mathrm{b}}$ one has
\bea\label{2}
F_{e0}(x_e,\rho) &=& \frac{x_e^2(1-\rho-x_e)^2}{(1-x_e)^3}[(1-x_e)(3-2x_e)+
\rho(3-x_e)]\\\label{3}
J_{e0}(x_e,\rho) &=& \frac{x_e^2(1-\rho-x_e)^2}{(1-x_e)^3}[(1-x_e)(1-2x_e)-
\rho(1+x_e)]\,.
\eea
The $1/m_{\mathrm{b}}^2$ nonperturbative corrections to these relations
have been
computed in \cite{Blok,MW} and the $\alpha_s$-order correction in
\cite{PRL,CJ}.
First we will neglect all corrections and use the free-quark decay results.
The energy of the electron in the laboratory frame is given by a boost
\bea\label{4}
E_\ell = \gamma(E_\ell^* + \beta p_\parallel^*)
\eea
where center-of-mass quantities are starred. With the help of this relation
the moments of the laboratory frame electron energy can be reduced to the
moments of the two functions in (\ref{1})
\bea\label{5}
F_{en} &=& \int_0^{1-\rho}\mbox{d}x_e\,x_e^n F_e(x_e,\rho)\\
\label{6}
J_{en} &=& \int_0^{1-\rho}\mbox{d}x_e\,x_e^n J_e(x_e,\rho)\,.
\eea

\begin{center}
\begin{tabular}{|c|c|c|}
\hline
 & &  \\[-0.3cm]
$n$ &$ \displaystyle \frac{\langle x_e^n\rangle}{\langle\gamma^n\rangle}$ &
$\displaystyle \frac{\langle x_\nu^n \rangle}{\langle\gamma^n\rangle}$
\\[0.4cm]
\hline
\hline
& & \\[-0.3cm]
$\displaystyle
1$ & $\displaystyle\frac{7-P}{10}-\rho\frac{19-7P}{10}$ &
$\displaystyle\frac{3+P}{5}(1-2\rho)$   \\[0.4cm]
\hline
& & \\[-0.3cm]
2 & $\displaystyle\frac{8(4-P)}{45}-\rho\frac{16(11-5P)}{45}$ &
$\displaystyle\frac{4(2+P)}{15}(1-4\rho)$ \\[0.4cm]
\hline
& & \\[-0.3cm]
3 & $\displaystyle\frac{2(3-P)}{7}-\rho\frac{2(25-13P)}{7}$ &
$\displaystyle\frac{4(5+3P)}{35}(1-6\rho)$ \\[0.4cm]
\hline
& & \\[-0.3cm]
4 & $\displaystyle\frac{8(5-2P)}{35}-\rho\frac{64(7-4P)}{35}$ &
$\displaystyle\frac{8(3+2P)}{35}(1-8\rho)$  \\[0.4cm]
\hline
& & \\[-0.3cm]
5 & $\displaystyle\frac{4(11-5P)}{27}-\rho\frac{20(31-19P)}{27}$ &
$\displaystyle\frac{32(7+5P)}{252}(1-10\rho)$ \\[0.4cm]
\hline
& & \\[-0.3cm]
6 & $\displaystyle\frac{128(2-P)}{105}-\rho\frac{256(17-11P)}{105}$ &
$\displaystyle\frac{32(4+3P)}{105}(1-12\rho)$ \\[0.4cm]
\hline
\end{tabular}
\end{center}
\begin{quote} {\bf Table 1.} Average values of the first six moments
of the electron (second column) and neutrino energy (third column)
in the laboratory frame, to first order in $\rho$.
\end{quote}

To first order in $\rho$ they are given by
\bea
F_{en} &=& \frac{6+n}{(3+n)(4+n)} -3\rho\frac{4+n}{3+n} + {\cal O}(\rho^2)
\\
J_{en} &=& -\frac{2+n}{(3+n)(4+n)}+3\rho\frac{2+n}{3+n} + {\cal O}(\rho^2)
\,.
\eea

 In Table 1 the moments of the reduced lepton energy in the laboratory
frame are listed as functions of $P$, to first order in $\rho$.
$\langle\beta\rangle$ has been taken equal to 1 in computing these
expressions, since the $\mathrm{b}$--quarks produced at the $\mathrm{Z}$
 pole are practically ultrarelativistic.
The qualitative behaviour of the $P$-dependence is just what one expects from
the general considerations mentioned at the beginning of this section: the
moments of the $x_e$-distribution grow as $P$ approaches the completely
polarized value $-1$. The relative increase with respect to the unpolarized
case grows with $n$, due to the fact that the higher moments
increasingly probe the region of high values of $x_e$, where the spectrum is
hardened in the polarized case.

   Unfortunately, the higher moments display also a larger sensitivity to
$\rho$, which is the most important source of uncertainty in these
calculations.
Previous works used a rather broad range of values 0.06-0.12
\cite{AM,M,BR} for $\rho$, which obtains when the
individual quark masses are varied independently within the
limits $m_{\mathrm{b}} = 4.8\pm 0.3$ GeV and $m_{\mathrm{c}} = 1.35\pm 0.15$
GeV.
It is worth noting that a simple use of the heavy mass expansion in the
HQET can be used to narrow down this interval.
This can be done by writing simultaneous heavy mass expansions
for the masses of the corresponding heavy mesons
\bea\label{9}
& &\frac14(m_B + 3m_{B^*}) = m_{\mathrm{b}}
+ \Lambda + \frac{\mu_\pi^2}{2m_{\mathrm{b}}} +
{\cal O}(1/m_{\mathrm{b}}^2)\\\label{10}
& &\frac14(m_D + 3m_{D^*}) = m_{\mathrm{c}} + \Lambda + \frac{\mu_\pi^2}
{2m_{\mathrm{c}}} +
{\cal O}(1/m_{\mathrm{c}}^2)\,.
\eea
The binding $\Lambda$ and the heavy quark kinetic energy $\mu_\pi^2$ are
still imprecisely known parameters. For example, in \cite{BGSUV} the bound
$\Lambda
> 500$ MeV was obtained and in \cite{hadr} correlated bounds on $\Lambda$
and $\mu_\pi^2$ are given, as well as the lower bound $\Lambda>410$ MeV.
Lattice calculations give much smaller results \cite{Lat} for
$\Lambda$.
We will use as a conservative number $\Lambda = 450\pm 50$ MeV.

 A QCD sum rule calculation \cite{BB} gave the value
$\mu_\pi^2 = (0.60\pm 0.10)$ GeV$^2$, in agreement with model-independent
bounds derived in \cite{BSUV,BGSUV}. We will use this central value but
double the error bars. On the left-hand side of (\ref{9},\ref{10}) we use
average masses over the respective isospin doublets, which gives \cite{PDG}
$\frac14(m_D + 3m_{D^*})=1973$ MeV and $\frac14(m_B + 3m_{B^*})=5279$ MeV.
Neglecting the terms of order $1/m^2$ and higher in (\ref{9}-\ref{10}) one
obtains $(m_{\mathrm{b}},\,m_{\mathrm{c}}) = (4.745~$GeV$,\,1.355~$GeV$)\div
(4.871~$GeV$,\,1.547~
$GeV) and correspondingly $\rho= 0.0815\div 0.1009$. These numbers are
similar to
those quoted in \cite{Voloshin}, i.e. $m_{\mathrm{b}}=4.80\pm 0.03$ GeV and
$m_{\mathrm{c}}=1.35\pm
0.05$ GeV (see also the references cited in \cite{Voloshin}).

  Returning to the analysis of the $\rho$-dependence of the moments of the
electron energy spectrum, we present in Table 2 the values of these moments
corresponding to the fully polarized $\Lambda_{\mathrm{b}}$ case ($P=-1$) and
in the
unpolarized case ($P=0$).
In computing these numbers the dependence on $\rho$ has been fully taken
into account, not only to first order as in Table 1.

\begin{center}
\begin{tabular}{|c|cc|cc|}
\hline
& & & &\\[-0.3cm]
\rule{0pt}{11.5pt}
& \multicolumn{2}{|c|}{
$\displaystyle \frac{\langle x_e^n\rangle}{\langle\gamma^n\rangle}$}
& \multicolumn{2}{c|}{
$\displaystyle \frac{\langle x_\nu^n\rangle}{\langle\gamma^n\rangle}$}
\\[0.4cm] \cline{2-5}
$n$ & $P=-1$  & $P=0$ & $P=-1$  & $P=0$ \\[0.4cm]
\hline
\hline
& & & &\\[-0.3cm]
$1$ & $\displaystyle 0.648^{-0.012}_{+0.013}$ &
$\displaystyle 0.583^{-0.010}_{+0.010}$ &
$\displaystyle 0.346^{-0.005}_{+0.005}$ &
$\displaystyle 0.519^{-0.007}_{+0.007}$ \\[0.4cm]
\hline
& & & &\\[-0.3cm]
$2$ & $\displaystyle 0.592^{-0.021}_{+0.023}$ &
$\displaystyle 0.496^{-0.016}_{+0.017}$ &
$\displaystyle 0.200^{-0.005}_{+0.006}$ &
$\displaystyle 0.400^{-0.011}_{+0.011}$ \\[0.4cm]
\hline
& & & &\\[-0.3cm]
$3$ & $\displaystyle 0.632^{-0.033}_{+0.035}$ &
$\displaystyle 0.503^{-0.024}_{+0.026}$ &
$\displaystyle 0.149^{-0.006}_{+0.006}$ &
$\displaystyle 0.373^{-0.015}_{+0.016}$ \\[0.4cm]
\hline
& & & &\\[-0.3cm]
$4$ & $\displaystyle 0.741^{-0.050}_{+0.054}$ &
$\displaystyle 0.566^{-0.035}_{+0.038}$ &
$\displaystyle 0.130^{-0.007}_{+0.007}$ &
$\displaystyle 0.391^{-0.021}_{+0.022}$ \\[0.4cm]
\hline
& & & &\\[-0.3cm]
$5$ & $\displaystyle 0.924^{-0.076}_{+0.084}$ &
$\displaystyle 0.684^{-0.052}_{+0.057}$ &
$\displaystyle 0.127^{-0.008}_{+0.009}$ &
$\displaystyle 0.444^{-0.029}_{+0.031}$ \\[0.4cm]
\hline
& & & &\\[-0.3cm]
$6$ & $\displaystyle 1.207^{-0.115}_{+0.129}$ &
$\displaystyle 0.870^{-0.078}_{+0.087}$ &
$\displaystyle 0.133^{-0.010}_{+0.011}$ &
$\displaystyle 0.533^{-0.041}_{+0.045}$ \\[0.4cm]
\hline
\end{tabular}
\end{center}
\begin{quote} {\bf Table 2.} Sensitivity to $P$ versus sensitivity to
$\rho$ in the electron and neutrino spectrum. The value
$\rho = 0.091\pm 0.010$ has been used (see the explanation in the
text).
\end{quote}

  The result is that, for the first moments, the error due to $\rho$ is
only marginally smaller than the sensitivity to $P$ \cite{AM,M}. With
growing $n$ both of
them increase such that for larger $n$ the uncertainty induced by $\rho$
is offset by the increase in the sensitivity to $P$ and the efficiency of
the variable becomes better. However, the additional uncertainty connected
with the fragmentation-dependent factor $\langle\gamma^n\rangle$, the need
of considering high moments in which the statistical errors become
appreciable and the
comparatively small efficiency make the use of the electron spectrum
little attractive as a polarization analyzer.

  Similar considerations can be made for the case of the inclusive neutrino
energy spectrum. This is described by a formula analogous to (\ref{1}) with
the corresponding functions $F_\nu$ and $J_\nu$ reducing to the well-known
free-quark decay expressions in the infinite mass limit
\bea
F_{\nu}(x_\nu,\rho) = J_{\nu}(x_\nu,\rho) =
\frac{6x_\nu^2(1-\rho-x_\nu)^2}{1-x_\nu}\,.
\eea
Their moments can be expanded to first order in $\rho$ with the result
\bea
F_{\nu n}=J_{\nu n} = \frac{6}{(n+3)(n+4)} - \frac{12\rho}{n+3} +
{\cal O}(\rho^2)\,.
\eea
The results for the first few moments of the neutrino reduced energy in the
laboratory frame are shown in the third column of Table 1. Their relative
sensitivity to $P$ and $\rho$ are given in the last two columns of Table 2.
One can see that the main conclusions drawn in the electron spectrum case can
be extended to the neutrino case.

   More satisfactory variables are obtained by taking the ratios of the
moments of the electron and neutrino spectrum. Numerically, they are much
more stable against
changes in $\rho$ because the electron and neutrino moments go in the same
direction under a variation in $\rho$. At the same time, the sensitivity
to $P$ is strongly enhanced \cite{BR} (see Table 3). Most importantly, the
boost-dependent factors $\langle\gamma^n\rangle$ cancel out in the ratios
$y_n$, which are therefore independent of details of the fragmentation
process \cite{BR}.

 The sensitivity to $P$ increases rapidly with $n$, from a relative change
of 60\% in $y_1$ as $P$ goes from 0 to 1 to over 450\% for $y_6$.
In \cite{BR} this relative change has been used as a measure for the
sensitivity of a given variable to the polarization. It is clear, however,
that the effect of the statistical errors on each particular variable has to
be taken into account as well. We propose instead as a more realistic measure
of the relative efficiency of two different variables, the ratio of the
errors on $P$ induced by each of them with identical statistical data.

\begin{center}
\begin{tabular}{|c|cc|cc|}
\hline
& & & &\\[-0.3cm]
& \multicolumn{2}{|c|}{
$\displaystyle y_n =
 \frac{\langle x_e^n\rangle}{\langle x_\nu^n\rangle}$}
& \multicolumn{2}{c|}{
$\frac{\displaystyle\delta P_n}{\displaystyle\delta P_1}$}
\\[0.4cm] \cline{2-5}
$n$ & $P=-1$  & $P=0$ & $P=-1$  & $P=0$ \\[0.4cm]
\hline
\hline
& & & &\\[-0.3cm]
$1$ & $\displaystyle 1.873^{-0.009}_{+0.010}$ &
$\displaystyle 1.124^{-0.003}_{+0.003}$ &
1.000 & 1.000\\[0.4cm]
\hline
& & & &\\[-0.3cm]
$2$ & $\displaystyle 2.957^{-0.026}_{+0.028}$ &
$\displaystyle 1.239^{-0.007}_{+0.007}$ &
$0.814\gamma_2$ &
$1.055\gamma_2$\\[0.4cm]
\hline
& & & &\\[-0.3cm]
$3$ & $\displaystyle 4.233^{-0.051}_{+0.054}$ &
$\displaystyle 1.347^{-0.010}_{+0.011}$ &
$0.893\gamma_3$ &
$1.352\gamma_3$\\[0.4cm]
\hline
& & & &\\[-0.3cm]
$4$ & $\displaystyle 5.684^{-0.085}_{+0.090}$ &
$\displaystyle 1.447^{-0.014}_{+0.015}$ &
$1.037\gamma_4$ &
$1.756\gamma_4$\\[0.4cm]
\hline
& & & &\\[-0.3cm]
$5$ & $\displaystyle 7.294^{-0.127}_{+0.135}$ &
$\displaystyle 1.542^{-0.018}_{+0.019}$ &
$1.219\gamma_5$ &
$2.253\gamma_5$\\[0.4cm]
\hline
& & & &\\[-0.3cm]
$6$ & $\displaystyle 9.053^{-0.177}_{+0.189}$ &
$\displaystyle 1.632^{-0.022}_{+0.024}$ &
$1.431\gamma_6$ &
$2.842\gamma_6$\\[0.4cm]
\hline
\end{tabular}
\end{center}
\begin{quote} {\bf Table 3.} Ratios of the moments of the electron and
neutrino energy spectra.
\end{quote}

  Explicitly, assuming that the variable $y_n$ can be measured with a
statistical error $\delta y_n$, this allows for a determination of $P$ with
an error
\bea
\delta P = \frac{\displaystyle\delta y_n}{\displaystyle\vert
\frac{\displaystyle dy_n}{\displaystyle dP}\vert}\,.
\eea

The error of $y_n$ is, in turn, given by
\bea
\delta y_n=\frac{ny_n}{\sqrt{N}}\sqrt{\frac{\langle E_e^{2(n-1)}\rangle}
{\langle E_e^n\rangle^2} (\delta E_e)^2 +
\frac{\langle E_\nu^{2(n-1)}\rangle} {\langle E_\nu^n\rangle^2}
(\delta E_\nu)^2}\,,
\eea
where $N$ is the number of events and $\delta E_{e,\nu}$ are the
statistical errors with which the electron and neutrino energies are
measured. Assuming that the systematic errors dominate over the statistical
ones, we will take $\delta E_e\simeq \delta E_\nu$. Using the data in Table 2
for the average energies one obtains the numbers shown in the last two
columns of Table 3 for the
ratio of the errors in $P$ induced by $y_n$ and $y_1$.
The boost-dependent factors $\gamma_n$
are given by
\bea
\gamma_n = \frac{\sqrt{\langle\gamma^{2(n-1)}\rangle}\langle\gamma\rangle}
{\langle\gamma^n\rangle}
\eea
and are of the order of the unity. For example, by using a Peterson
ansatz for the fragmentation function of the form $f(z)=\frac{1}{4z(1-
\frac{1}{z}-\frac{\epsilon}{1-z})^2}$ with $\epsilon=0.016$
the following average values are obtained: $\langle\gamma\rangle=6.244$,
$\gamma_2=0.981$, $\gamma_3=0.989$
and $\gamma_4=1.009$.

One can see that, as far as the statistical errors are
concerned, $y_2$ allows an
extraction of $P$ with an error at least 20\% smaller than $y_1$ (note
that $\gamma_2<1$ independent of the fragmentation model adopted)
in the fully polarized case ($P=-1$).
 In the vicinity of $P=0$ the two variables are equally
sensitive to statistical errors.

   Incorporating the nonperturbative corrections amounts to replacing the
free-quark decay moments (\ref{5},\ref{6}) according to $F_1\to F_1$,
$J_1\to
(1+\epsilon_{\mathrm{b}})J_1$, $F_2\to (1+\frac53 K_{\mathrm{b}})F_2$, $J_2
\to (1+\epsilon_{\mathrm{b}}+
\frac53 K_{\mathrm{b}})J_2$ \cite{MW}. The nonperturbative matrix elements
appearing
in these relations have the form $K_{\mathrm{b}}=\mu_\pi^2/2m_{\mathrm{b}}^2$
and $\epsilon_{\mathrm{b}}=
\mu_s^2/m_{\mathrm{b}}^2$. We will use the value $\mu_s^2=-\mu_\pi^2/3$
suggested by
a model-independent bound \cite{bound} and $\mu_\pi^2=0.6\pm 0.2$ GeV$^2$
as already explained above.

   The analytic expressions for the radiative corrections can be found in
\cite{CJ}. We have computed the corresponding corrections to the moments of
the electron and neutrino spectrum
numerically. The value of the strong coupling $\alpha_s$ has been varied
between 0.1 and 0.3. The results, including the errors associated with the
variation of the parameters involved, are presented in Table 4.

\begin{center}
\begin{tabular}{|c|c|c|c|c||c|}
\hline
& & & & & \\[-0.3cm]
variable & central & $\rho$ & $\mu_\pi^2$ & $\alpha_s$ & Total \\[0.4cm]
\hline
\hline
& & & & & \\[-0.3cm]
$y_1(P=0)$ & 1.131 & $^{+0.003}_{-0.003}$ & $^{+0.000}_{-0.000}$ &
$^{+0.004}_{-0.003}$ & $^{+0.007}_{-0.007}$\\[0.4cm]
\hline
& & & & & \\[-0.3cm]
$y_1(P=-1)$ & 1.901 & $^{+0.011}_{-0.010}$ & $^{+0.004}_{-0.004}$ &
$^{+0.011}_{-0.010}$ & $^{+0.027}_{-0.023}$\\[0.4cm]
\hline
& & & & &\\[-0.3cm]
$R_{y_1}$ & 1.681 & $^{+0.005}_{-0.005}$ & $^{+0.003}_{-0.003}$ &
$^{+0.004}_{-0.003}$ & $^{+0.013}_{-0.011}$\\[0.4cm]
\hline
& & & & &\\[-0.3cm]
$y_2(P=0)$ & 1.253 & $^{+0.007}_{-0.007}$ & $^{+0.000}_{-0.000}$ &
$^{+0.007}_{-0.008}$ & $^{+0.016}_{-0.014}$\\[0.4cm]
\hline
& & & & &\\[-0.3cm]
$y_2(P=-1)$ & 3.039 & $^{+0.039}_{-0.029}$ & $^{+0.011}_{-0.011}$ &
$^{+0.031}_{-0.028}$ & $^{+0.076}_{-0.065}$\\[0.4cm]
\hline
& & & & &\\[-0.3cm]
$R_{y_2}$ & 2.425 & $^{+0.011}_{-0.010}$ & $^{+0.009}_{-0.009}$ &
$^{+0.009}_{-0.008}$ & $^{+0.030}_{-0.026}$\\[0.4cm]
\hline
\end{tabular}
\end{center}
\begin{quote} {\bf Table 4.} The contributions of the different sources
of theoretical errors to $y_1$ and $y_2$. $R_{y_n}(P)\equiv y_n(P)/y_n(0)$
are the ratios $y_n$ normalized at 1 at $P=0$.
\end{quote}

\begin{figure}[hhh]
 \begin{center}
 \mbox{\epsfig{file=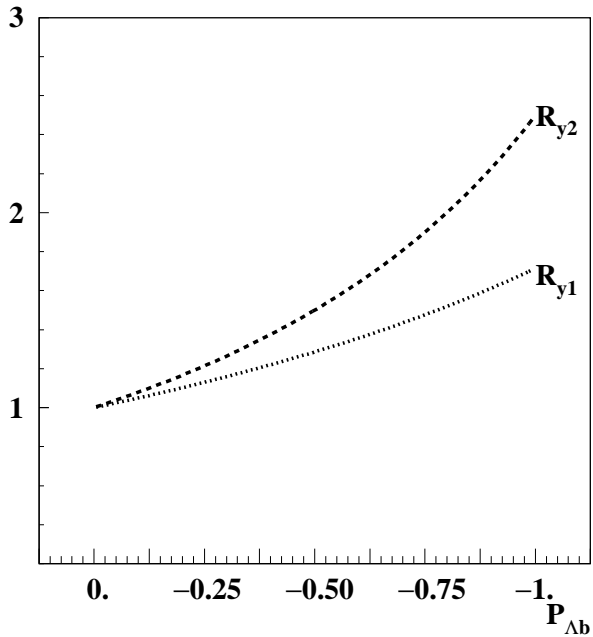,width=10cm}}
 \end{center}
 \caption{
The $R_{y_1}$ and
$R_{y_2}$ variables  versus the $\Lambda_{\mathrm{b}}$
polarization for $\rho=0.09$.}
\label{ry1ry2}
\end{figure}

   The systematic theoretical errors can be further reduced by
considering the ratios $R_{y_n}(P)=y_n(P)/y_n(P=0)$ \cite{PPE}. Since these
ratios differ from $y_n$ only by a constant, the arguments
following after Eq.(15) concerning the statistical errors apply to them
without changes.
 Assuming zero statistical errors, the total theoretical uncertainty
of $R_{y_1}$ is reflected into an error of 2.3\% in $P$ at $P=-1$,
compared with about 4.3\% when considering $y_1$ at the same point.
In a similar way, $R_{y_2}$ induces an uncertainty in $P$ of 2.0\% at
$P=-1$, compared to an error decreasing from 3.9\% at the same point to
3.3\% at $P=0$ for $y_2$. In the Figure~\ref{ry1ry2},
 the $R_{y_1}$ and
$R_{y_2}$ variables are plotted as a function of the $\Lambda_{\mathrm{b}}$
polarization for $\rho=0.09$.

\thispagestyle{plain}

  In conclusion, $y_2$ and $R_{y_2}$ could prove to be useful as variables
for measuring the $\Lambda_{\mathrm{b}}$ polarization in addition to $y_1$
(respectively $R_{y_1}$), due to their reduced sensitivity to statistical
errors.

\section{Other polarization-dependent variables}

 In this section we discuss a few other variables which can be used to
measure the polarization of the $\Lambda_{\mathrm{b}}$ baryons. For example,
one could consider the average ratio of the electron and neutrino energies.
In terms of rest-frame quantities this can be written as
\bea\label{yDdef}
y_D = \langle\frac{E_e}{E_\nu}\rangle =
\frac{1}{\Gamma_t}\int\mbox{d}\Gamma\,\frac{E_e^*+\beta p_{e\parallel}^*}
{E_\nu^*+\beta p_{\nu\parallel}^*}\,.
\label{yDirect}
\eea
We will keep $\beta$ fixed for the moment and only average over it at the
very end. Let us compute the average (\ref{yDdef}) in the free-quark decay
approximation.
The differential decay rate $\mbox{d}\Gamma$ is given by
\bea
\frac{1}{\Gamma_t}\mbox{d}\Gamma = \frac{3}{2\pi f(\rho)}
x_e^2 x_\nu^2 (1+P\cos\theta_\nu) \mbox{d}x_\nu \mbox{d}\cos\theta_\nu
\mbox{d}\cos\theta_{e\nu} \mbox{d}\phi\,,
\eea
where
\bea
x_e = \frac{1-\rho-x_\nu}{1+\frac{x_\nu}{2}(\cos\theta_{e\nu}-1)}
\eea
and $f(\rho)=1-8\rho+8\rho^3-\rho^4-12\rho^2\log\rho$. $\theta_{e\nu}$
is the angle between the electron and neutrino moments in the rest frame of
the decay and $\phi$ denotes the angle between the decay plane and the
$(\vec p_\nu,\,\vec p_\Lambda)$ plane in the same reference frame. After
integrating over all variables
in (\ref{yDirect}) the following result is obtained
\bea\label{yD}
y_D &=& \frac{6}{f(\rho)}\left\{-\frac{1}{12}(1-\rho)(1-11\rho-47\rho^2-3
\rho^3) + \rho^2(3+2\rho)\log\rho\right\}\\
&+&\, \frac{1}{f(\rho)}\left(2\frac{P}{\beta}+\frac{\beta-P}{\beta^2}
\log\frac{1+\beta}{1-\beta}\right)\left[(1-\rho)(1-8\rho-17\rho^2)-
6\rho^2(3+\rho)\log\rho\right]\,.\nonumber
\eea

\begin{figure}[hhh]
 \begin{center}
 \mbox{\epsfig{file=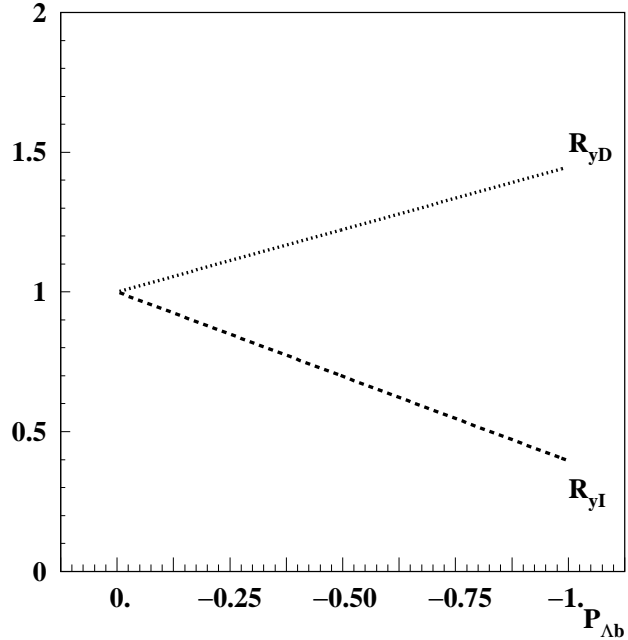,width=10cm}}
 \end{center}
 \caption{
$R_{y_{D}}$ and $R_{y_{I}}$ as functions of the $\Lambda_{\mathrm{b}}$
polarization.}
\label{rydryi}
\end{figure}

Since $\beta$ is very close to 1, it can be replaced everywhere with this
value, except in the argument of the logarithm, which can be written as
\bea
\log\frac{1+\beta}{1-\beta} = 2\log 2\gamma + {\cal O}(\beta-1)\,.
\eea
With this approximation, (\ref{yD}) becomes
\bea\label{yDfinal}
y_D &=& \frac{1}{2f(\rho)}\left\{-\frac{1}{12}(1-\rho)(1-11\rho-47\rho^2
-3\rho^3) + \rho^2(3+2\rho)\log\rho\right\}\\
&+&\,\frac{2}{f(\rho)}\left(P + (1-P)\langle\log 2\gamma\rangle
\right)\left[(1-\rho)(1-8\rho-17\rho^2)-
6\rho^2(3+\rho)\log\rho\right]\,.\nonumber
\eea


   The fragmentation-dependent quantity $\langle\log 2\gamma\rangle$ can be
eliminated by considering also the average of the inverse ratio
\bea\label{yInverse}
y_I = \langle\frac{E_\nu}{E_e}\rangle =
\frac{1}{\Gamma_t}\int\mbox{d}\Gamma\,\frac{E_\nu^*+\beta p_{\nu\parallel}^*}
{E_e^*+\beta p_{e\parallel}^*}\,.
\eea
This can be computed in a similar way to $y_D$ with the somewhat lengthy
result
\bea
y_I &=&
\frac{1}{f(\rho)}\left\{-\frac{1}{12}(1-\rho)(3-29\rho-65\rho^2+31\rho^3)
+\rho^2(6-\rho^2)\log\rho\right\}\nonumber\\
&+&
\frac{2\langle\log 2\gamma\rangle}{f(\rho)}\left\{\frac13(1-\rho)(2-13\rho-
4\rho^2+3\rho^3)-2\rho^2(3-\rho)\log\rho\right\}\nonumber\\
&+&\label{yIfinal}
\frac{2}{f(\rho)}P\left\{-\frac16(1-\rho)^2(1+10\rho+\rho^2)-\rho(1-\rho^2)
\log\rho\right\}\\
&+&
\frac{2\langle\log 2\gamma\rangle}{f(\rho)}P\left\{\frac13(1-\rho)(1-8\rho-
17\rho^2)-2\rho^2(3+\rho)\log\rho\right\}\nonumber\\
&-&
\frac{2}{f(\rho)}\langle\frac{\log 2\gamma}{\gamma^2}\rangle
P\left\{\frac16(1-\rho)(1-5\rho+13\rho^2+3\rho^3)+2\rho^3\log\rho\right\}
\nonumber\,.
\eea
To a good approximation, the last term in (\ref{yIfinal}) can be neglected
because of the
additional suppression by a factor of $1/\gamma^2$. An estimate performed
with the
help of the Peterson fragmentation function used in the preceding section
gives $\langle\log 2\gamma\rangle\simeq 2.497$ and $\langle(\log 2\gamma)
/\gamma^2\rangle \simeq 0.064$.

\par
The  variable $y_D$  should be used carefully; due to detector smearing
the neutrino energy can be close to zero and create overflows. The solution to
this problem is to do a cut on neutrino energy around zero. This cut together
with other analysis cuts ({\it e.g.} lepton identification cuts on the lepton
momentum at $3\,GeV$ and on its transverse momentum at $1\,GeV/c$)
will bias  the values of  $y_D$ and $y_I$.
 The normalizations $R_{y_D} = y_D(P)/y_D(0)$ and $R_{y_I} = y_I(P)/y_I(0)$
 can be used  to eliminate the acceptance and
reconstruction effects (see Figure~\ref{rydryi}).
A Monte Carlo study shows that
the relative errors obtained in a sample of N events
are the following
$$
\begin{array}{ccc}
 & & \\
 &P_{\Lambda_{\mathrm{b}}} = 0.&P_{\Lambda_{\mathrm{b}}} = -1. \\
 & & \\
\displaystyle \frac{\sigma_{R_{y_D}}}{R_{y_D}}
&\displaystyle \frac{2.4}{\sqrt{N}}
&\displaystyle \frac{2.0}{\sqrt{N}} \\
 & & \\
\displaystyle \frac{\sigma_{R{y_I}}}{R{y_I}}
&\displaystyle \frac{2.4}{\sqrt{N}}
&\displaystyle \frac{2.8}{\sqrt{N}} \\
 & & \\
\end{array}
$$

  A simultaneous fit to (\ref{yDfinal}) and (\ref{yIfinal}) will give the
polarization $P$. The fragmentation variable $\langle \log 2 \gamma \rangle$
can be eliminated between $R_{y_D}$ and $R_{y_I}$
 and the polarization value extracted
as a function of both neglecting the term proportional to
$\langle\frac{\log 2\gamma}{\gamma^2}\rangle$.

\par
  The fragmentation-dependence of the variables $y_D$ and $y_I$
(\ref{yDfinal},\ref{yIfinal}) can be traced back to the fact that the
energy ratios in (\ref{yDirect}) and respectively (\ref{yInverse}) are
themselves not
Lorentz-invariant under boosts along the direction of motion of the
$\Lambda_{\mathrm{b}}$. One can construct boost-invariant distributions and
averages (hence insensitive to the details of the fragmentation process)
by considering quantities which are explicitly invariant under such
Lorentz-transformations.

\begin{figure}[hhh]
 \begin{center}
 \mbox{\epsfig{file=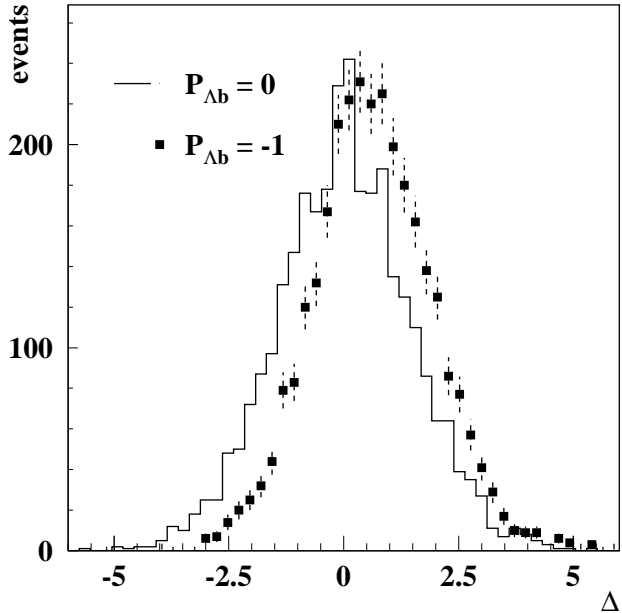,width=10cm}}
 \end{center}
 \caption{The  difference between charged lepton  and neutrino
rapidities $ \Delta $ for
two values of the $\Lambda_{\mathrm{b}}$ polarization
$P_{\Lambda_{\mathrm{b}}}=-1.$ and  $P_{\Lambda_{\mathrm{b}}}=0.$
}
\label{rap}
\end{figure}

  One such quantity, which turns out to be also sensitive to
$\Lambda_{\mathrm{b}}$ polarization, is the difference of the rapidities of the
electron and of the neutrino along the boost direction
\bea\label{Delta}
\Delta = \zeta_e - \zeta_\nu\,.
\eea
In general the rapidity of a particle is defined as $\zeta=\frac12\log
\frac{E+q}{E-q}$ with $E$ its energy and $q$ the longitudinal momentum.
For a massless particle the rapidity can be also expressed through the
angle $\theta$ between its direction of motion and the boost axis,
$\tanh\zeta = \cos\theta$.
We recall that in the polarized case ($P=-1$), the electrons tend to be
emitted preferentially along the boost direction and the neutrinos in the
opposite direction. This means that the distribution in $\Delta$ should be
shifted towards positive values for a polarized $\Lambda_{\mathrm{b}}$
compared with the unpolarized case (see Figure~\ref{rap}).

   The simplest way to obtain the decay distribution differential in
$\Delta$ is to parametrize the final state of the
system from the very beginning
in terms of the set of coordinates $(\zeta,r,\phi)$ for each
particle: the rapidity $\zeta$, the transverse momentum to the boost axis
$r$ and the azimuthal angle $\phi$.
This gives
\bea
\mbox{d}\Gamma &=& \frac{G_F^2|V_{cb}|^2}{2(2\pi)^5}r_e^2
(m_{\mathrm{b}}^2-m_{\mathrm{c}}^2-2m_{\mathrm{b}}r_e\cosh\zeta_e)^2\nonumber\\
& &\times\,
\left[ m_{\mathrm{b}}^2\cosh(\zeta_e+\zeta_\nu)+m_{\mathrm{c}}^2\cosh
(\zeta_e-\zeta_\nu) +
(m_{\mathrm{b}}^2-m_{\mathrm{c}}^2)\cos(\phi_e-\phi_\nu)\right]\\
& &\times\,(\cosh\zeta_\nu+P\sinh\zeta_\nu)
\frac{\mbox{d}r_e\mbox{d}\zeta_e\mbox{d}\zeta_\nu\mbox{d}\phi_e\mbox{d}
\phi_\nu}{\left\{ m_{\mathrm{b}}\cosh\zeta_\nu+r_e[\cos(\phi_e-\phi_\nu)-
\cosh(\zeta_e-\zeta_\nu)]\right\}^4}\nonumber\,.
\eea
The electron transverse momentum $r_e$ takes values from 0 to $m_{\mathrm{b}}
(1-\rho)/
(2\cosh\zeta_e)$. It is therefore helpful to introduce a reduced transverse
electron momentum $r$ defined as $r=2r_e\cosh\zeta_e/m_{\mathrm{b}}$
which will take values between 0 and $1-\rho$.

\begin{figure}[hhh]
 \begin{center}
 \mbox{\epsfig{file=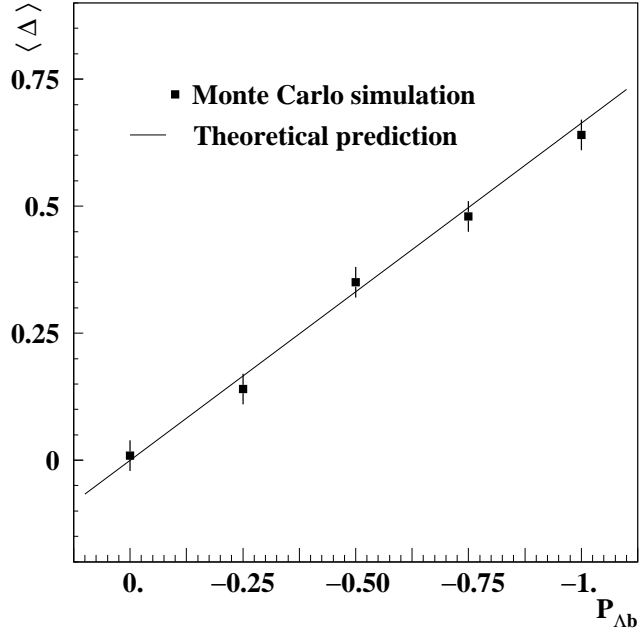,width=10cm}}
 \end{center}
 \caption{
 $\langle \Delta \rangle $ variable as a function of
 the $\Lambda_{\mathrm{b}}$ polarization  simulated for five different values
$P_{\Lambda_{\mathrm{b}}}=0.0,-0.25,-0.5,-0.75 \; \mathrm{and} \;-1.0$
(points) and the theoretical prediction (continous line). The Monte Carlo
samples contain 3000 events and the quark mass ratio has been taken
$\rho=0.09$.}
\label{averap}
\end{figure}

After integrating over $\phi_{e,\nu}$ and transforming from $(\zeta_e,
\zeta_\nu)$ to $\zeta_+ = \zeta_e+\zeta_\nu$ and $\Delta$ (\ref{Delta}),
the following result is obtained
\bea
\mbox{d}\Gamma &=& \frac{G_F^2|V_{cb}|^2}{8(2\pi)^3}m_{\mathrm{b}}^5\,
r(1-\rho-r)^2
\left\{(1-\rho)\frac{r^2+2(\Sigma-r\cosh\Delta)^2}{\sqrt{R^5}}\right.
\nonumber\\
& &\left. \quad-\, (1-\rho-r)\Sigma
\frac{[\Sigma-r\cosh\Delta][3r^2+2(\Sigma-r\cosh\Delta)^2]}{\sqrt{R^7}}
\right\}\nonumber\\
& &\times\left[\cosh\zeta_++\cosh\Delta + P\left(\sinh\zeta_+-\sinh\Delta
\right)\right]\mbox{d}r \mbox{d}\Delta\mbox{d}\zeta_+\,,\label{triple}
\eea
where $\Sigma=\cosh\zeta_++\cosh\Delta$ and
$R = r^2\sinh^2\Delta - 2r\Sigma\cosh\Delta + \Sigma^2$.

 The exact result obtained after integrating over $r$ and $\zeta_+$ is
very complicated, so we only give in analytic form the simpler expressions
corresponding to the case of a massless final quark $\rho=0$. In this limit
the decay distribution has the form
\bea
\frac{\mbox{d}\Gamma}{\mbox{d}\Delta} &=&
\frac{G_F^2|V_{cb}|^2}{4(2\pi)^3}m_{\mathrm{b}}^5\,
\left(f(\Delta) - P\sinh\Delta\, g(\Delta)\right)\,,
\eea
where both $f(\Delta)$ and $g(\Delta)$ are even functions of their
argument. For positive values of $\Delta$ they are given by
\bea
f(\Delta)&=&-I(\Delta)c\frac{36s^4+180s^2+175}{2s^8}+
\left(\log(1+e^\Delta)-\frac{\Delta}{2}\right)
\frac{4s^6+218s^4+715s^2+525}{3s^8}\nonumber\\
&-&\,\frac{34s^6+(827-132c)s^4+10(232-127c)s^2+1575(1-c)}{18s^8}
\eea
and
\bea
g(\Delta) &=& -I(\Delta)\frac{32s^4+130s^2+105}{2s^8}
+\left(\log(1+e^\Delta)-\frac{\Delta}{2}\right)
c\frac{4s^4+180s^2+315}{3s^8}\nonumber\\
&+&\,\frac{4(11-3c)s^4+5(64-43c)s^2+315(1-c)}{6s^8}\,.
\eea
We have denoted in these expressions
\bea
I(\Delta) = \frac{1}{\sinh\Delta}\left(-2\mbox{Li}_2(-e^\Delta)-\frac12
\Delta^2-\frac{\pi^2}{6}\right)
\eea
and $c=\cosh\Delta$, $s=\sinh\Delta$.

\begin{center}
\begin{tabular}{|c|c|c|}
\hline
$\rho$ & $C_1$ & $C_3$ \\
\hline
\hline
0.000 & 0.667 & 4.089   \\
\hline
0.081 & 0.633 & 3.695 \\
\hline
0.091 & 0.630 & 3.662 \\
\hline
0.101 & 0.627 & 3.631 \\
\hline
\end{tabular}
\end{center}
\begin{quote} {\bf Table 5.} Average values of the difference of the
electron and neutrino rapidities $\Delta=\zeta_e-\zeta_\nu$, and of its
cube $\Delta^3$, at $P=-1$, for a few values of $\rho$.
\end{quote}

   The singularity of $f(\Delta)$ and $g(\Delta)$ at $\Delta=0$ is only
apparent; at this point these functions have power series expansions of the
form
\bea
f(\Delta) &=& \left(\frac{62}{945}-\frac{8}{315}\log 2\right) -
\Delta^2 \left(-\frac{91}{2970}+\frac{32}{495}\log 2\right)\nonumber\\
& &-\, \Delta^4\left(\frac{285209}{3243240}-\frac{17632}{135135}\log 2\right)
+ {\cal O}(\Delta^6)\\
g(\Delta) &=& \left(\frac{73}{315}-\frac{32}{105}\log 2\right) -
\Delta^2 \left(\frac{655}{2079}-\frac{304}{693}\log 2\right)\nonumber\\
& & +\,\Delta^4\left(\frac{25279}{117936}-\frac{748}{2457}\log 2\right)
+ {\cal O}(\Delta^6)\,.
\eea

  As a measure of the polarization one can take the average value of an odd
power of $\Delta$. These are proportional to $P$ and have the form
\bea
\langle\Delta^n\rangle = -C_n P
\eea
with $C_n$ positive numbers. The first two coefficients $C_n$ ($n=1,3$),
obtained by numerically integrating (\ref{triple}), are shown in Table 5
for a few values of $\rho$. In Figure~\ref{averap} a Monte Carlo simulation
was used to estimate  $\langle\Delta\rangle$  as a function of polarization.

   With the given error on $\rho$, $\langle\Delta\rangle$ allows an
extraction of the
polarization with a theoretical uncertainty of the order of 2.1 \%
in the fully polarized case (and proportional to $P$ for $|P|<1$),
whereas $\langle\Delta^3\rangle$ gives a slightly larger
error of 2.3 \%. The statistical errors on the $\Lambda_{\mathrm{b}}$
polarization value are estimated in a Monte Carlo
simulation to be $0.9/\sqrt{N}$ with $N$ the number of events.
 While less precise than the variables discussed in
section 2, nevertheless we believe that the methods presented in this section
might prove to be useful as well, even if for cross-checking purposes only.

\acknowledgements

  It is a pleasure to thank G. Bonvicini for a critical reading of the
manuscript and for interesting comments. We are grateful to A.Czarnecki for
making available to us a FORTRAN
library containing the results of Ref.\cite{CJ} on radiative corrections
to polarized quark decay. One of us (D.P.) has greatly benefited from
many discussions with Bas Tausk. J.G.K. is supported in part by the
Bundesministerium f\"ur Forschung und Technologie, FRG, under contract
06MZ566, and D.P. ackowledges a grant from the Deutsche Physikalische
Gesellschaft, FRG.

\appendix
\section*{}

   We present here, for completeness, a few details concerning the radiative
corrections to the first moments of the electron and neutrino spectrum.
The corrections to the moments of the functions $F_{e,\nu}(x)$ and
$J_{e,\nu}(x)$ (describing the rest-frame decay distributions) have the
form
\bea
F_{en} &=& F_{en}^{(FQD)} - \frac{4\alpha_s}{\pi}\int_0^{1-\rho}\mbox{d}
x_e\, x_e^nf_1^-(x_e)\,,\,\,
J_{en} = J_{en}^{(FQD)} - \frac{4\alpha_s}{\pi}\int_0^{1-\rho}\mbox{d}
x_e\, x_e^nj_1^-(x_e)\,,\nonumber\\
F_{\nu n} &=& F_{\nu n}^{(FQD)} - \frac{4\alpha_s}{\pi}\int_0^{1-\rho}
\mbox{d}x_\nu\, x_\nu^nf_1^+(x_\nu)\,,\,\,
J_{\nu n} = J_{\nu n}^{(FQD)} - \frac{4\alpha_s}{\pi}\int_0^{1-\rho}
\mbox{d}x_\nu\, x_\nu^nj_1^+(x_\nu)\,.\nonumber
\eea

The polarization-dependent variables $y$ and $y_2$ are written in terms of
these moments as
\bea
y = \frac{3F_{e1} + J_{e1}P}{3F_{\nu1} + J_{\nu1}P}\,,\qquad
y_2 = \frac{2F_{e2} + J_{e2}P}{2F_{\nu2} + J_{\nu2}P}\,.
\eea

\begin{center}
\begin{tabular}{|c|c|c|}
\hline
$\rho$ & $F_{e0}^{(FQD)}$ & $f_0^-$ \\
\hline
\hline
$0.0815$ & $0.2760$ & $0.1173$   \\
\hline
$0.0912$ & $0.2577$ & $0.1074$ \\
\hline
$0.1009$ & $0.2406$ & $0.0984$ \\
\hline
\end{tabular}
\end{center}
\begin{quote} {\bf Table A1.} The zeroth moment of the function $F_e(x_e)$
appearing in the unpolarized $b$-quark decay rate and its associated
radiative correction.
\end{quote}

The functions $f_1^\pm(x)$ and $j_1^\pm(x)$ are defined in Eq.(16) and
thereafter of Ref.~\cite{CJ}.

\begin{center}
\begin{tabular}{|c|c|c|c|c|c|c|c|c|}
\hline
$\rho$ & $F_{e1}^{(FQD)}$ & $f_1^-$ & $J_{e1}^{(FQD)}$ & $j_1^-$ &
$F_{\nu 1}^{(FQD)}$ & $f_1^+$ & $J_{\nu 1}^{(FQD)}$ & $j_1^+$ \\
\hline
\hline
$0.0815$ & $0.1638$ & $0.0701$ & $-0.0556$ & $-0.0152$ & $0.1453$ &
$0.0651$ & $0.1453$ & $0.0659$ \\
\hline
$0.0912$ & $0.1503$ & $0.0630$ & $-0.0498$ & $-0.0137$ & $0.1337$ &
$0.0588$ & $0.1337$ & $0.0595$ \\
\hline
$0.1009$ & $0.1380$ & $0.0566$ & $-0.0447$ & $-0.0124$ & $0.1231$ &
$0.0531$ & $0.1231$ & $0.0537$ \\
\hline
\end{tabular}
\end{center}
\begin{quote} {\bf Table A2.} The $n=1$ moments and their radiative
corrections.
\end{quote}

In Table A1 we show the $n=0$ moment of $F_e(x_e)$ and its radiative
correction, which are needed for the normalization of the absolute values
of the higher moments. However, they cancel when considering the ratios of
moments $y_n$.

The values of the free-quark decay moments
$F_n^{(FQD)}$ and $J_n^{(FQD)}$ and the moments of $f_1^\pm(x)$,
$j_1^\pm(x)$ (denoted as $f_n^\pm$ and $j_n^\pm$) for $n=1$ and $n=2$
are listed in the tables A2 and A3.

\begin{center}
\begin{tabular}{|c|c|c|c|c|c|c|c|c|}
\hline
$\rho$ & $F_{e2}^{(FQD)}$ & $f_2^-$ & $J_{e2}^{(FQD)}$ & $j_2^-$ &
$F_{\nu 2}^{(FQD)}$ & $f_2^+$ & $J_{\nu 2}^{(FQD)}$ & $j_2^+$ \\
\hline
\hline
$0.0815$ & $0.1062$ & $0.0461$ & $-0.0420$ & $-0.0121$ & $0.0852$ &
$0.0403$ & $0.0852$ & $0.0407$ \\
\hline
$0.0912$ & $0.0960$ & $0.0407$ & $-0.0370$ & $-0.0106$ & $0.0774$ &
$0.0359$ & $0.0774$ & $0.0362$ \\
\hline
$0.1009$ & $0.0866$ & $0.0359$ & $-0.0327$ & $-0.0094$ & $0.0702$ &
$0.0320$ & $0.0702$ & $0.0323$ \\
\hline
\end{tabular}
\end{center}
\begin{quote} {\bf Table A3.} The $n=2$ moments and their radiative
corrections.
\end{quote}

{}From these tables one can see that the radiative corrections change the
$n=1$ moments by 10-17\% (for $\alpha_s=0.3$). The corresponding effect
in the $n=2$ moments is somewhat larger, of 11-18\%.



\begin{references}
\bibitem{1} J.G.K\"orner, A.Pilaftsis and M.Tung, {\em Z.Phys.}{\bf C63}
   575 (1994).
\bibitem{2} M.Tung, {\em Phys.Rev.}{\bf D52} 1353 (1995).
\bibitem{3} S.Groote, J.G.K\"orner and M.Tung, MZ-TH-95-09, July 1995,
   hep-ph/9507222
\bibitem{FP} A.Falk and M.Peskin, {\em Phys.Rev.}{\bf D49} 3320 (1994).
\bibitem{CKPS} F.Close, J.G.K\"orner, R.J.N.Phillips and D.J.Summers,
   {\em J.Phys.}{\bf G18} 1716 (1992).
\bibitem{PPE} The ALEPH Collaboration, {\em Measurement of the
   $\Lambda_{\mathrm{b}}$
   polarization in Z decays}, CERN PPE/95-156
\bibitem{BR} G.Bonvicini and L.Randall, {\em Phys.Rev.Lett.}{\bf 73}
   392 (1994).
\bibitem{AM} B.Mele and G.Altarelli, {\em Phys.Lett.}{\bf B299} 345 (1993).
\bibitem{M} B.Mele, {\em Mod.Phys.Lett.}{\bf A9} 1239 (1994).
\bibitem{KK} J.K.Kim and Y.G.Kim, {\em Phys.Rev.}{\bf 52} 5352 (1995).
\bibitem{CGG} J.Chay, H.Georgi and B.Grinstein, {\em Phys.Lett.}{\bf B247}
   399 (1990).
\bibitem{Blok} B.Blok, L.Koyrakh, M.Shifman and A.I.Vainshtein,
   {\em Phys.Rev.}{\bf D49} 3356 (1994).
\bibitem{MW} A.Manohar and M.Wise, {\em Phys.Rev.}{\bf D49} 1310 (1994).
\bibitem{tr1} J.H.K\"uhn, A.Reiter and P.M.Zerwas, {\em Nucl.Phys.}
   {\bf B272} 560 (1986).
\bibitem{tr2} S.Groote and J.G.K\"orner, MZ-TH/95-17, hep-ph/9508399
\bibitem{PRL} A.Czarnecki, M.Jezabek, J.G.K\"orner and J.H.K\"uhn,
   {\em Phys.Rev.Lett.}{\bf 73} 384 (1994).
\bibitem{CJ} A.Czarnecki and M.Jezabek, {\em Nucl.Phys.}{\bf B427} 3
   (1994).
\bibitem{hadr} A.Falk, M.Luke and M.J.Savage, {\em Hadron spectra for
   semileptonic decay}, UTPT-95-11, July 1995.
\bibitem{BB} P.Ball and V.Braun, {\em Phys.Rev.}{\bf D49} 2472 (1994).
\bibitem{Lat} M.Crisafulli, V.Gimenez, G.Martinelli and C.T.Sachrajda,
   {\em First lattice calculation of the B meson binding and kinetic
   energies}, CERN-TH-7521-94, June 1995, hep-ph/9506210
\bibitem{PDG} Particle Data Group, {\em Phys.Rev.}{\bf D50} 1175 (1994).
\bibitem{Voloshin} M.B.Voloshin, {\em Phys.Rev.}{\bf D51} 4934 (1995).
\bibitem{BSUV} I.Bigi, M.Shifman, N.G.Uraltsev and
   A.Vainshtein, {\em Phys.Rev.}{\bf D52} 196 (1995).
\bibitem{BGSUV} I.Bigi, A.G.Grozin, M.Shifman, N.G.Uraltsev and
   A.Vainshtein,\\ {\em Phys.Lett.}{\bf B339} 160 (1994).
\bibitem{bound} J.G.K\"orner and D.Pirjol, {\em Phys.Lett.}{\bf B334}
   399 (1994).
\end{references}
\end{document}